\pdfoutput=1

\documentclass[11pt,twoside,a4paper,cmspaper,final,collab]{cms-tdr}
\def\svnVersion{49480M}\def\cmsCernNo{2011-035}\def\cmsCernDate{\today}\def\cmsMessage{Submitted to the Journal of High Energy Physics}

\begin{document}\cmsNoteHeader{EWK-10-013}

\hyphenation{had-ron-i-za-tion}
\hyphenation{cal-or-i-me-ter}
\hyphenation{de-vices}
\RCS$Revision: 48858 $
\RCS$HeadURL: svn+ssh://alverson@svn.cern.ch/reps/tdr2/papers/EWK-10-013/trunk/EWK-10-013.tex $
\RCS$Id: EWK-10-013.tex 48858 2011-04-06 13:04:13Z monicava $
\cmsNoteHeader{EWK-10-013} 
\title{Measurement of the Inclusive $\mathrm{Z}$ Cross~Section via Decays to
Tau Pairs in $\Pp\Pp$ Collisions at $\sqrt{s}=7$~TeV}

\address[cern]{CERN}
\author[cern]{The CMS Collaboration}

\date{\today}

\abstract{
A measurement of inclusive $\mathrm{Z}\rightarrow\tau^+\tau^-$ production in pp collisions
is presented, in the final states $\mu$+hadrons, $\mathrm{e}$+hadrons, $\mathrm{e}$+$\mu$,
and $\mu$+$\mu$. The data sample corresponds to an integrated luminosity
of $36~\mathrm{pb}^{-1}$ collected with the CMS detector at the LHC.
The measured cross section is
$\sigma \left (\mathrm{p}\mathrm{p}\to \mathrm{Z} X \right) \times {\cal{B}} \left(\mathrm{Z} \to \tau^+\tau^- \right) = 1.00 \pm 0.05\,\mathrm{(stat.)} \pm 0.08\,\mathrm{(syst.)} \pm 0.04\,\mathrm{(lumi.)}\,\mathrm{nb}$,
which is in good agreement with the next-to-next-to-leading order
QCD prediction and with previous measurements in the
$\mathrm{Z}\rightarrow \mathrm{e}^+\mathrm{e}^-$ and $\mu^+\mu^-$ channels.
The reconstruction efficiency for hadronic $\tau$ decays is determined
with a precision of $7\%$.
}

\hypersetup{%
pdfauthor={CMS Collaboration},%
pdftitle={Measurement of the Inclusive Z Cross~Section via Decays to
Tau Pairs in pp Collisions at sqrt(s)=7 TeV},%
pdfsubject={CMS},%
pdfkeywords={CMS, physics}}

\maketitle 

\newcommand{\Ztautau}    {\mathrm{Z}\rightarrow\Pgt^+\Pgt^-}
\newcommand{\tauhad}{\ensuremath{\Pgt_{\rm had}}}
\newcommand{\taue}{\ensuremath{\Pgt_\Pe}}
\newcommand{\taumu}{\ensuremath{\Pgt_\Pgm}}
\newcommand{\Irel}{I_{\mathrm{rel}}^{\mathrm{PF}}}
\newcommand{\Irelemu}{I_{\mathrm{rel}}}
\newcommand{\PZ}    {\mathrm{Z}}

\renewcommand{\MET}{\ensuremath{{E\!\!\!/}_{\mathrm{T}}}\xspace}

\newcommand{\MT}{\ensuremath{{M}_{\mathrm{T}}}\xspace}

\def\ERROR#1#2{ \ensuremath{ \pm #1\, (\textrm{#2}) }  }
\def\RESA#1#2#3#4{ \ensuremath{ #1 \ERROR{#2}{stat.} \ERROR{#3}{syst.} \ERROR{#4}{lumi.} \, \mathrm{nb} }}
\def\RESB#1#2#3#4#5{ \ensuremath{ #1 \ERROR{#2}{stat.} \ERROR{#3}{syst.} \ERROR{#4}{lumi.} \ERROR{#5}{\Pgt-\mathrm{ID}} \, \mathrm{nb} }}
\def\SIGBR#1#2{  \ensuremath{ \sigma \left( \Pp\Pp \to #1 X \right) \times {\cal{B}} \left( #1 \to #2 \right) } }

\section{Introduction} \label{sec:introduction}
\par
The measurement of the production
cross section for $\Pp\Pp \rightarrow \PZ X$ with 
$\Ztautau$ constitutes an important physics benchmark at the Large Hadron Collider (LHC).
The $\Pgt$ lepton can decay either into purely leptonic final states
($\Pgt\rightarrow \Pe\Pagne\Pgngt$ denoted as ``$\taue$''
or $\Pgt\rightarrow \Pgm\Pagngm\Pgngt$ denoted as ``$\taumu$'')
or into hadronic final states denoted by ``$\tauhad$''
consisting of a hadronic system and a~$\nu_{\Pgt}$.
Constrained by the $\Pgt$ mass, the hadronic system is
characterized by a low particle multiplicity and
a highly collimated jet which allows a $\tauhad$ signal
to be separated from the large QCD jet backgrounds.
The validation of the $\tauhad$ signal
is essential in searches for new physics based
on $\Pgt$ leptons, such as Higgs boson
decays to $\Pgt^+\Pgt^-$~\cite{Higgs_guide}.
The Compact Muon Solenoid (CMS) Collaboration recently reported a search for the Higgs boson
in this channel~\cite{ourHtautaupaper}.
Tau leptons can also be important signatures for searches of 
supersymmetry, extra dimensions, and extra gauge bosons~\cite{PTDR2}.

The $\Ztautau$ production cross section has been previously measured
in proton-antiproton collisions by the CDF and D0 Collaborations~\cite{Abulencia:2007iy,Abazov:2008ff}.
In this study, $\Ztautau$ events
in the $\taumu\tauhad$, $\taue\tauhad$,
$\taue\taumu$, and $\taumu\taumu$
final states are selected from a sample of $\sqrt{s} = 7$~TeV proton-proton collision data
recorded by the CMS experiment
at the LHC. The data sample corresponds to
an integrated luminosity of $36 \pm 1~{\mathrm{pb}}^{-1}$.
The results are compared to previous measurements made
in the $\Pep\Pem$ and $\Pgmp\Pgmm$ final states~\cite{Khachatryan:2010xn},
providing a validation of
$\tauhad$ reconstruction and
identification~\cite{CMS_PFT_08-001,PAS_PFT-09-001,PAS_TAU-11-001} and a direct measurement of the tau selection efficiency.

\section{CMS Detector}\label{sec:dectector}
The central feature of the CMS apparatus 
is a superconducting solenoid, of 6~m internal diameter, providing 
a magnetic field of 3.8~T. Within the field volume are the silicon pixel 
and strip tracker, the crystal electromagnetic calorimeter (ECAL) 
and the brass/scintil\-lator hadron calorimeter (HCAL). 
Muons are measured in gas-ionization detectors embedded 
in the steel return yoke.

CMS uses a right-handed coordinate system, with the 
origin at the nominal interaction point, the $x$ axis 
pointing to the centre of the LHC, the $y$ axis pointing 
up perpendicular to the LHC plane, and the $z$ axis 
along the counterclockwise-beam direction. The polar angle  
$\theta$ is measured from the positive $z$ axis and the 
azimuthal angle $\phi$ is measured in the $xy$ plane.
Variables used in this article are
the pseudorapidity,
$\eta \equiv -\ln[\tan(\theta/2)]$, and the transverse
momentum, $\PT = \sqrt{p_x^2 + p_y^2}$.

The ECAL is designed to have both excellent energy resolution
and high granularity, which are crucial for reconstructing
electrons and photons produced in $\Pgt$~decays.
The ECAL is constructed with projective lead tungstate 
crystals in two pseudorapidity
regions: the barrel ($|\eta| < 1.479$) and the endcap
($1.479 < |\eta| < 3$). 
The transition regions between the barrel and the endcaps, $ 1.444 < |\eta| < 1.567$, 
are not used for electron reconstruction. 
In the barrel region, the crystals
are $25.8~X_0$ long, where $X_0$ is the radiation length, and conform to a granularity of
$\Delta\eta\times\Delta\phi = 0.0174\times0.0174$.
The endcap region is instrumented with a lead--silicon-strip 
preshower detector consisting of two orthogonal strip
detectors with a strip pitch of 1.9~mm.  One plane is
at a depth of $2X_0$ and the other at $3X_0$.
The ECAL has an energy resolution of better than $0.5\%$ for unconverted photons with transverse energies above 100\GeV. The energy resolution is $3\%$ or better for the range of electron energies relevant for this analysis.
The HCAL barrel and endcap regions cover the range $|\eta| < 3$ 
and are subdivided
into towers with a segmentation of 
$\Delta\eta\times\Delta\phi = 0.087\times0.087$,
corresponding to $5\times 5$ ECAL crystals in the
barrel region. The HCAL forward region extends the calorimetry 
to  $|\eta| < 5$.

The inner tracker measures charged particle tracks
within the range $|\eta| < 2.5$.  It consists of 
$1\, 440$ silicon pixel and $15\, 148$ silicon strip detector modules
and provides an impact parameter resolution of 
$\sim$\,15~$\Pgm$m and a transverse momentum 
resolution of about $1.5\%$ for 
100\GeV particles. 
 
The muon barrel region is covered
by drift tubes and the endcap regions by cathode 
strip chambers.  In both regions resistive plate
chambers provide additional coordinate and timing
information.  Muons can be reconstructed in the
range $|\eta| < 2.4$, with a typical $\PT$
resolution of $1\%$ for $\PT \approx 40$\GeV.

A more detailed description of CMS can be 
found in~\cite{JINST}.

\section{Lepton Reconstruction and Identification} \label{sec:leptonid}

Muons produced by $\Pgt$ decays in the $\Ztautau$ process
are reconstructed in the tracker and muon chambers~\cite{CMS-PAS-MUO-10-002}. 
Quality cuts, based on the minimum number of hits in the silicon tracker, pixel detector, 
and muon chambers, are applied to suppress backgrounds
from punch-throughs and decays in flight.

Electrons are reconstructed by combining tracks produced by the 
Gaussian Sum Filter algorithm with ECAL 
clusters~\cite{CMS-PAS-EGM-10-004}.  Requirements are imposed that 
distinguish prompt electrons from charged pions mimicking electron
signatures, and from electrons produced by photon conversions.  

The CMS particle flow (PF) algorithm~\cite{PAS_PFT-09-001} is used
to form a mutually exclusive collection of reconstructed particles 
(muons, electrons, photons, and charged and neutral hadrons)
by combining tracks and calorimeter clusters.
These reconstructed particles are used to build composite objects such as $\Pgt$'s and 
jets, and to measure the missing transverse energy $\MET$.

Electrons and muons from $\Pgt$ decays are expected to be 
isolated in the detector, while leptons from 
heavy-flavour ($\mathrm{c}$ and $\mathrm{b}$) decays and decays in flight are expected to be found inside jets.
A measure of isolation is used to discriminate the signal from
the QCD multijet background, based on the charged hadrons, photons, and
neutral hadrons 
falling within
a cone $\Delta R = \sqrt{\Delta\eta^2+\Delta\phi^2} = 0.4$ around the lepton momentum direction.  
In the $\taumu\taumu$ final state, a cone of $\Delta R = 0.3$ is used.  
A sum of the $\PT$ for each particle type is made  
for charged hadrons with
$\PT > 0.5$\GeV and for photons and neutral hadrons
with $\PT > 1$\GeV; photons and neutral hadrons are 
excluded from the sum if they fall within inner cones
of $\Delta R = 0.05$ and $\Delta R = 0.08$, respectively.
The relative isolation variable is
$
 \Irel =  \Sigma \left(
  \PT^{{\mathrm{charged}}} +
  \PT^{{\mathrm{photon}}} + 
  \PT^{{\mathrm{neutral}}} 
         \right)  / \PT^{\ell}
$
, where $\PT^{{\mathrm{charged}}}$, $\PT^{{\mathrm{photon}}}$,  
and $\PT^{{\mathrm{neutral}}}$  refer to the charged hadrons,
photons, and neutral hadrons in the sum, respectively, and $\PT^{\ell}$ refers to the $\PT$ of the lepton $\ell = \Pe ,\Pgm$.
For muons, it is required that $\Irel < 0.1$, while for electrons
it is required that $\Irel < 0.08$ in the barrel and $\Irel < 0.04$
in the endcaps.    A similar formula is used for the 
$\taue\taumu$ final state but with the isolation quantities 
based directly on tracker and calorimeter information, calculated in a cone of $\Delta R = 0.3$.  
In this case, it is required that $\Irelemu < 0.15$ for muons and $\Irelemu < 0.1$ for electrons.

The $\tauhad$ identification algorithm used in this measurement 
is known as the Hadrons Plus Strips algorithm~\cite{PAS_TAU-11-001}, 
which starts from a high-$\PT$ charged hadron and combines it 
with other nearby charged or neutral hadrons to reconstruct
$\Pgt$ decay modes. The identification of $\Pgpz$ mesons is 
enhanced by clustering electrons and photons in 
''strips'' along the bending plane to take into account 
possible broadening of calorimeter signatures by 
photon conversions. To reduce the contamination from QCD jets, 
the $\tauhad$-candidate isolation is calculated
in a cone of $\Delta R = 0.5$ around the reconstructed
$\Pgt$-momentum direction.  
It is requested that there be no charged hadrons with $\PT>1.0$\GeV and 
no photons with $\ET > 1.5$\GeV in the isolation cone, other than the $\Pgt$ constituents.

\section{Event Selection} \label{sec:channels-selection}
The events preselected for this analysis in the $\taumu\tauhad$, $\taue\taumu$, and $\taumu\taumu$ final states are required to pass the single-muon Level-1 (L1) trigger with a $\PT$ threshold of 7\GeV and the single muon High Level Trigger (HLT)~\cite{HLT}, with a threshold varying from 9\GeV to 15\GeV, depending on the instantaneous luminosity.
Events in the $\taue\tauhad$ final state are selected using the single-electron L1 trigger with a threshold of 8\GeV in the transverse energy, and  a single-electron HLT trigger with a threshold of 12\GeV at higher instantaneous luminosity.   
At the end of the 2010 data taking period, a combined $\Pe$+$\Pgt$ trigger was used in order to keep the rate low enough without a further increase of the electron threshold. The trigger has the same L1 requirements as the electron trigger, but 
the HLT requires the presence of an electron of transverse energy larger than 12\GeV and a hadronic tau decay with $\PT$ larger than 15\GeV, tagged with a simplified version of the tau reconstruction algorithm with a less restrictive selection than the one used in the offline analysis.

\begin{figure}[!h]
\begin{center}
\includegraphics[scale=0.35]{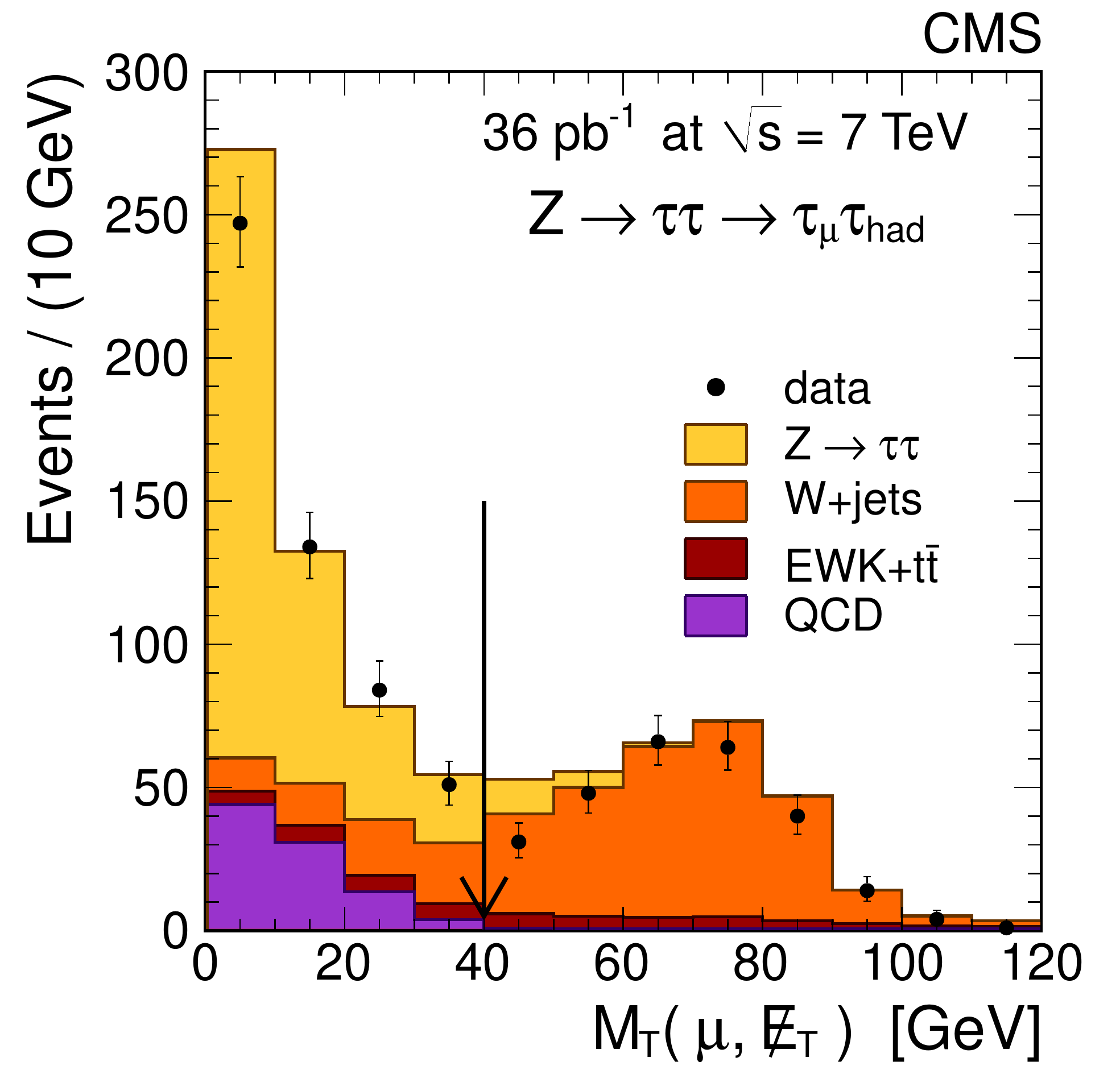}
\includegraphics[scale=0.35]{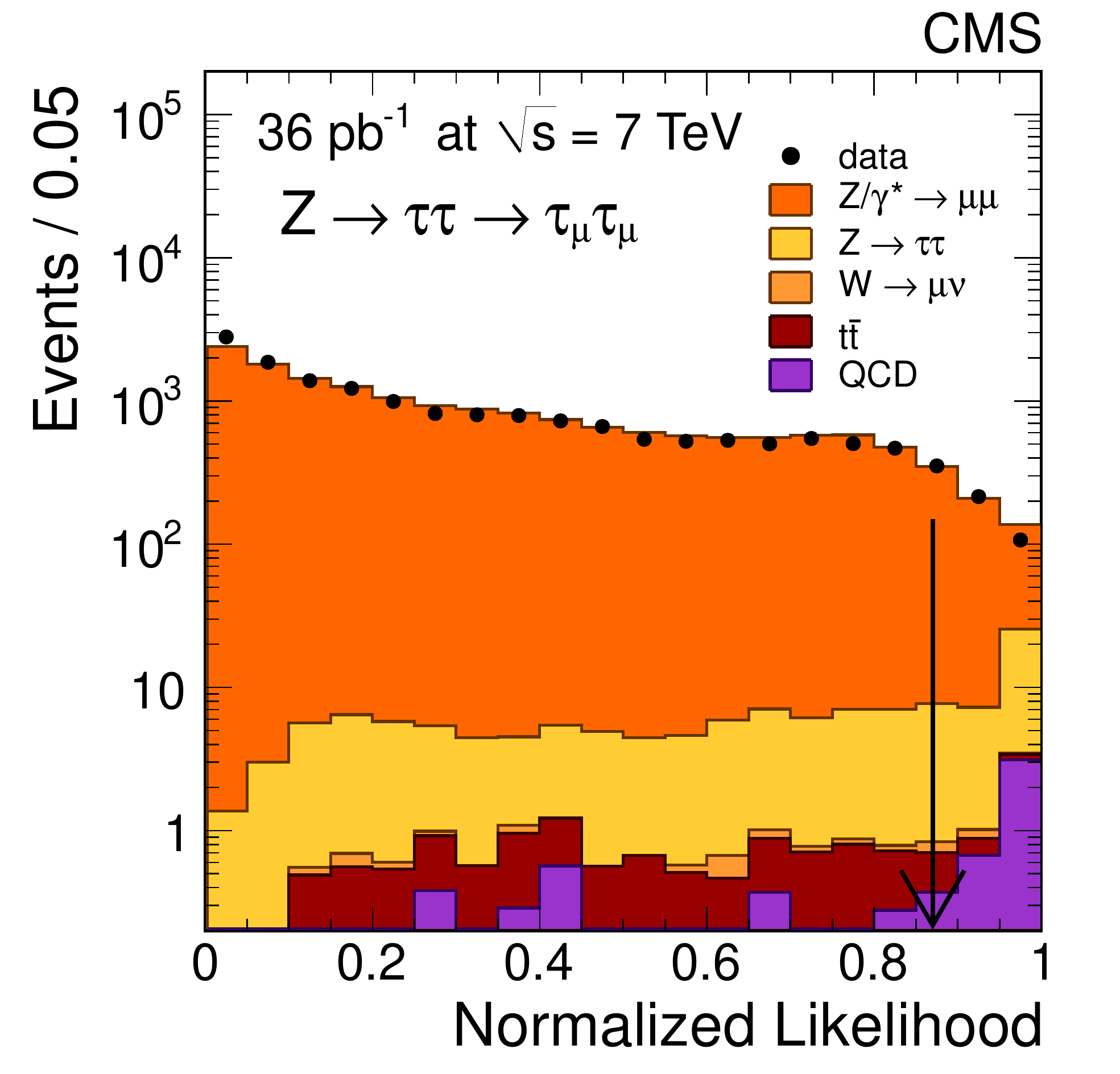}
\caption{Muon+$\MET$ transverse mass (left) for the $\taumu\tauhad$ final state. Likelihood discriminant (right) for the  
$\taumu\taumu$ final state. The arrows show the final selection criteria applied. 
}
\label{figure:control-plot}
\end{center}
\end{figure}

Offline event selection starts with the requirement of a
well-defined primary vertex~\cite{CMS-PAS-TRK-10-005}.  
For the $\taumu\tauhad$ and $\taue\tauhad$ final states, 
one isolated muon or electron is required with $\PT > 15\GeV$   
and $|\eta| < 2.1$.  The associated $\tauhad$ must be 
oppositely charged, with $\PT > 20\GeV$ and $|\eta| < 2.3$. 
In order to reject events coming from the $\PW$+jets background, 
the transverse mass of the lepton and the $\MET$, $\MT(\ell,\MET) = \sqrt{ 2 \PT^{\ell} \MET \cdot \left ( 1 - \cos {\Delta \phi} \right ) }$, is required to be less than  $40\GeV$, 
where 
$ \Delta \phi$ is the difference in azimuthal angle between the lepton and $\MET$ vectors.
The $\MT$ distribution and the selection requirement applied are illustrated in Fig.~\ref{figure:control-plot} (left)  for the $\taumu\tauhad$ final state.

For the  $\taue\taumu$ final state, one isolated muon 
with $|\eta| < 2.1$ and one oppositely charged isolated 
electron with $|\eta| < 2.4$ are required, both with  
$\PT > 15\GeV$. Further background suppression is achieved 
by requiring $\MT(\Pgm,\MET) < 50\GeV$ and $\MT(\Pe,\MET) < 50\GeV$.

For the $\taumu\taumu$ final state, events with two oppositely charged 
isolated muons with $|\eta| < 2.1$ are selected if one satisfies
$\PT > 19\GeV$ and the other $\PT > 10\GeV$.  The
requirement $\MET < 50\GeV$ suppresses $\ttbar$ and
$\PW$+jets backgrounds, and a requirement on the azimuthal
angle difference between the muons, $\Delta\phi_{\Pgm\Pgm} > 2$,    
rejects QCD background events, for muons originating from the same quarkonia decay 
or from a decay chain of heavy-flavour hadrons.
Effective suppression of the Drell--Yan background is achieved 
with a multivariate likelihood ratio technique. 
For each event, the relative probabilities to belong to 
two event classes ($\Ztautau\rightarrow\taumu\taumu$ and 
$\PZ /\gamma^{*}\rightarrow\Pgmp\Pgmm$) are computed, exploiting the
following set of discriminating variables: the ratio of the 
transverse momentum 
of the dimuon system to the scalar sum of the momenta of the two muons;  
the significance of the distance of closest 
approach (DCA) between the two muon helix tracks; the pseudorapidity of the dimuon system; and the 
azimuthal angle between the positive muon momentum and  $\MET$. A normalized likelihood discriminant is computed and is shown in Fig.~\ref{figure:control-plot} (right).
Events with a likelihood discriminant value larger than 0.87 are kept 
in the final sample. 

For the plots in Fig.~\ref{figure:control-plot}, the QCD multijet, $\PW$ and diboson backgrounds were  
simulated with {\sc PYTHIA}~\cite{Sjostrand:2006za}, the 
Drell--Yan signal and background with the next-to-leading order (NLO) Monte Carlo generator 
{\sc POWHEG}~\cite{Alioli:2008gx,Nason:2004rx,Frixione:2007vw}, and the top samples with Madgraph~\cite{Maltoni:2002qb}.
The tau decays were performed with Tauola~\cite{Jadach:1993hs}.
 The samples were normalized using the cross section at next-to-next-to-leading order (NNLO) for Drell--Yan and $\PW$, at leading order (LO) for QCD, and NLO for the $\ttbar$ sample. 
The EWK label in the figure refers to the $\PZ \rightarrow\Pep\Pem, \Pgmp\Pgmm$, and diboson backgrounds. 

\section{Background Estimation}\label{sec:background}
The $\Ztautau$ signal is established using the visible mass, which is 
the reconstructed mass of the $\ell\tauhad$ system in the  $\Pgt_{\ell}\tauhad$ 
final states, and the dilepton invariant mass 
for the $\taue\taumu$ and $\taumu\taumu$ final states.
Due to the presence of neutrinos in the final state, the 
$\Ztautau$ visible mass peak extends across the range
$30$--$100$\GeV, which is considerably broader than
the $\PZ $ resonance itself.  The backgrounds generally 
span the same mass range, so an accurate determination
of the signal yield requires effective background estimation
techniques.   The main background sources are  
QCD multijet processes, $\PW$+jets, and $\PZ \rightarrow\ell^+\ell^-$, 
with small contributions from top-quark decays 
and dibosons. All backgrounds are measured in control regions where 
their contributions are enhanced and extrapolated to the signal selection region 
using selection 
efficiencies determined either from data or the Monte Carlo simulation. 
In the $\Pgt_{\ell}\tauhad$ final state, the QCD multijet background is 
estimated using samples of same-sign (SS) and opposite-sign (OS) 
events from data, for which the electron or muon isolation requirement is 
inverted; the QCD background estimate is based on the ratio of
OS to SS yields.  
The visible mass distribution of the SS events is used to describe  
the background  in the signal region. The $\PW$ contribution 
is extracted from the region $\MT(\ell,\MET) > 60$\GeV,
where it dominates the sample. In the $\taue\taumu$ final state,  
the background contributions are expected to be small and are 
estimated from the simulation. The large-transverse-mass region 
is used to check the estimated background contributions from 
$W$, diboson, and $\ttbar$ backgrounds. For the  $\taumu\taumu$ 
final state, the Drell--Yan muon-pair production events are selected 
with a reduced likelihood without including the muon DCA significance,   
and the resulting muon DCA significance distribution is fitted with signal and background shapes. The 
QCD multijet background estimate is obtained from a sample of SS dimuon events.

The numbers of selected events and the expected  
background contributions are summarized in 
Table~\ref{table:background_tot}.  The statistical and systematic uncertainties in the methods used are also given, where the uncertainties
are added in quadrature.

\begin{table}[!h]
\caption{Numbers of expected background events and  number of data events passing all the selection criteria in the four final states. The uncertainties shown include the statistical and systematic uncertainties added in quadrature.}
\vspace*{0.2cm}
\label{table:background_tot}
\centering
\begin{tabular}{|l|c|c|c|c|}
\hline
 & $\taumu\tauhad$  & $\taue\tauhad$ &  $\taue\taumu$ &  $\taumu\taumu$  \\ 
 & &  &  &  ($M_{\Pgm\Pgm} < 70$\GeV) \\ 
\hline\hline
$\PZ \rightarrow\ell^+\ell^-$, jet misidentified as $\Pgt$ & 6.4 $\pm$ 2.4 & 15.0 $\pm$ 6.2 & \multicolumn{2}{c|}{-} \\
\hline
$\PZ \rightarrow\ell^+\ell^-$, lepton misidentified as $\Pgt$ & 12.9 $\pm$ 3.5 & 109 $\pm$ 28 & 2.4 $\pm$ 0.3 & 20.1 $\pm$ 1.3 \\ \cline{1-3} \cline{4-5}
\hline
$\ttbar$ & 6.0 $\pm$ 3.0 & 2.6 $\pm$ 1.3 & 7.1 $\pm$ 1.3 & 0.15 $\pm$ 0.03 \\
\hline
$\PW\rightarrow\ell\nu$ & 54.9 $\pm$ 4.8 &  30.6 $\pm$ 3.1 & &  \\ \cline{1-3} 
$\PW\rightarrow\Pgt\nu$ & 14.7 $\pm$ 1.3 & 7.0 $\pm$ 0.7 & 1.5 $\pm$ 0.5 &  2.5 $\pm$ 2.5  \\ \cline{1-3}
QCD multijet & 132 $\pm$ 14 & 181 $\pm$ 23 &  &  \\ \cline{1-3} 
\hline
$\PW\PW/\PW\PZ/\PZ\PZ$ & 1.6 $\pm$ 0.8 & 0.8 $\pm$ 0.4 & 3.0  $\pm$ 0.4  & - \\ 
\hline
\hline
Total background & 228 $\pm$ 16 & 346 $\pm$ 37 & 14.0 $\pm$ 1.8 & 22.8 $\pm$ 2.8 \\
\hline
\hline
Total data & 517 & 540 & 101 & 58 \\
\hline
\end{tabular}
\end{table}

\section{Systematic Uncertainties} \label{sec:systematics}
The efficiencies for electron and muon reconstruction, identification, and isolation,  
as well as the trigger efficiencies are obtained from data. 
Correction factors for the values extracted from the simulation are determined  
using the method described in Ref. ~\cite{Khachatryan:2010xn} ({\it tag-and-probe} method). 
The measured efficiencies have a small dependence on $\PT$ and cover the full range of $\PT$ used in the analysis. The uncertainties on the correction factors are in the range of $0.2$--$1.1\%$.

A similar technique is used to estimate the hadronic tau identification efficiency.
A data sample of taus is selected using $\Ztautau \rightarrow \taumu \tauhad$ events. 
The events are
preselected without applying the full tau identification but only kinematic cuts and a set of
requirements to suppress the background from $\PZ  \rightarrow \Pgmp \Pgmm$, $\PW$, and QCD events.
The efficiency is then calculated as a ratio of the number of events
that pass the tau identification requirement to the number of all preselected events.
The total uncertainty of the measurement arises from the statistical uncertainty of the sample 
and the systematic uncertainties related to the understanding of the backgrounds 
in the preselection sample and amounts to $23\%$~\cite{PAS_TAU-11-001}. 

To estimate the efficiency of the $\MT$ selection and the likelihood selection efficiency  
for the $\taumu\taumu$ final state, an embedded sample is used 
where the muons in a $\PZ \rightarrow\Pgmp\Pgmm$ data sample are 
replaced by simulated tau decays with the original muon momentum. The estimated uncertainties amount to $2\%$.

To estimate the effect of the energy-scale uncertainties on the acceptance, the 
energy of all reconstructed objects (electrons, muons, taus, and jets)
is varied within their respective uncertainty.
After each independent shift, the missing 
transverse energy is recalculated and the event selection is repeated.
The event  yield is compared to the nominal value and the relative difference is quoted 
as the systematic~uncertainty. The systematic uncertainties are in the range from $1\%$ to $3.5\%$.

To obtain the acceptance corrections, the D6T and Z2 {\sc PYTHIA} tunes~\cite{MCtunes} 
were used in the simulation. The effect of the use of different tunes on the final extracted cross section
is smaller than $1\%$ and is not included in the systematic uncertainties.

The main source of theoretical uncertainty in the calculation of the experimental acceptance comes from the parton distribution functions (PDFs). The central acceptance value is obtained with the CT10 PDF~\cite{Guzzi:2011sv}. The uncertainty is estimated using the error sets of the PDFs: CT10, MSTW2008NLO~\cite{Martin:2009iq}, and NNPDF2.0~\cite{Ball:2010de}, and amounts to~$2\%$.

The experimental and theoretical uncertainties are 
summarized in Table~\ref{table:systematics}.

\begin{table}[!h]
\caption{Summary of the sources of systematic uncertainties and their estimated effect on the measured $\Ztautau$ cross section.}
\begin{center}
\begin{tabular}{|l||c|c|c|c|} 
\hline
Source                                           & $\taumu\tauhad$  & $\taue\tauhad$ &  $\taue\taumu$ &  $\taumu\taumu$  \\ 
\hline
\hline
Trigger                                          & $0.2\%$              & $3\%$ & $0.2\%$ & $0.3\%$ \\ \hline
Lepton identification and isolation              & $1.0\%$            & $1.1\%$ & $1\%$ & $1\%$  \\ \hline
$\tauhad$ identification                         & \multicolumn{2}{c|}{$23\%$} & \multicolumn{2}{c|}{-}     \\ \hline
Efficiency of $M_T$ selection            & \multicolumn{2}{c|}{$2\%$}  & \multicolumn{2}{c|}{-}    \\ \hline
Likelihood selection efficiency                  &  \multicolumn{3}{c|}{-}     & $2\%$                      \\ \hline
Acceptance due to $\tauhad$ energy scale, $3\%$              & \multicolumn{2}{c|}{$3.5\%$} &  \multicolumn{2}{c|}{-}    \\ \hline
Acceptance due to e energy scale, $2\%$           &       -        &      $1.6\%$ & $1.6\%$ & -  \\ \hline
Acceptance due to $\mu$ momentum scale, $1\%$      & $1\%$            &  - & $1\%$ &  $2\%$  \\ \hline
Luminosity                                       & \multicolumn{4}{c|}{ $4\%$ }  \\ \hline      
Parton distribution functions                    & \multicolumn{4}{c|}{ $2\%$ } 
  \\ \hline      
\end{tabular}
\end{center}
\label{table:systematics}
\end{table}

\section{Cross Section Measurement}  \label{sec:results}
\par

The cross section is obtained for each
final state with the following formula: 
\begin{equation}
\sigma(\Pp\Pp\rightarrow \PZ X)\times {\cal{B}}(\Ztautau)=\frac{N}{{\cal{A}}\,\epsilon\, 
{{\cal{B}}^{'}} \, {\cal{L}}}\,\,\, ,
\label{eq:crosssection}
\end{equation}
where $N$ is the number of extracted signal events, $\cal{A}$ is the acceptance
of signal events, $\epsilon$ is the signal selection efficiency, ${{\cal{B}}^{'}}$ 
is the branching fraction of the $\Pgt$-decay mode considered~\cite{Nakamura:2010zzi}, and $\cal{L}$ is the 
integrated luminosity~\cite{LUMI}.

\begin{figure}[!h]
\begin{center}
\includegraphics[scale=0.35]{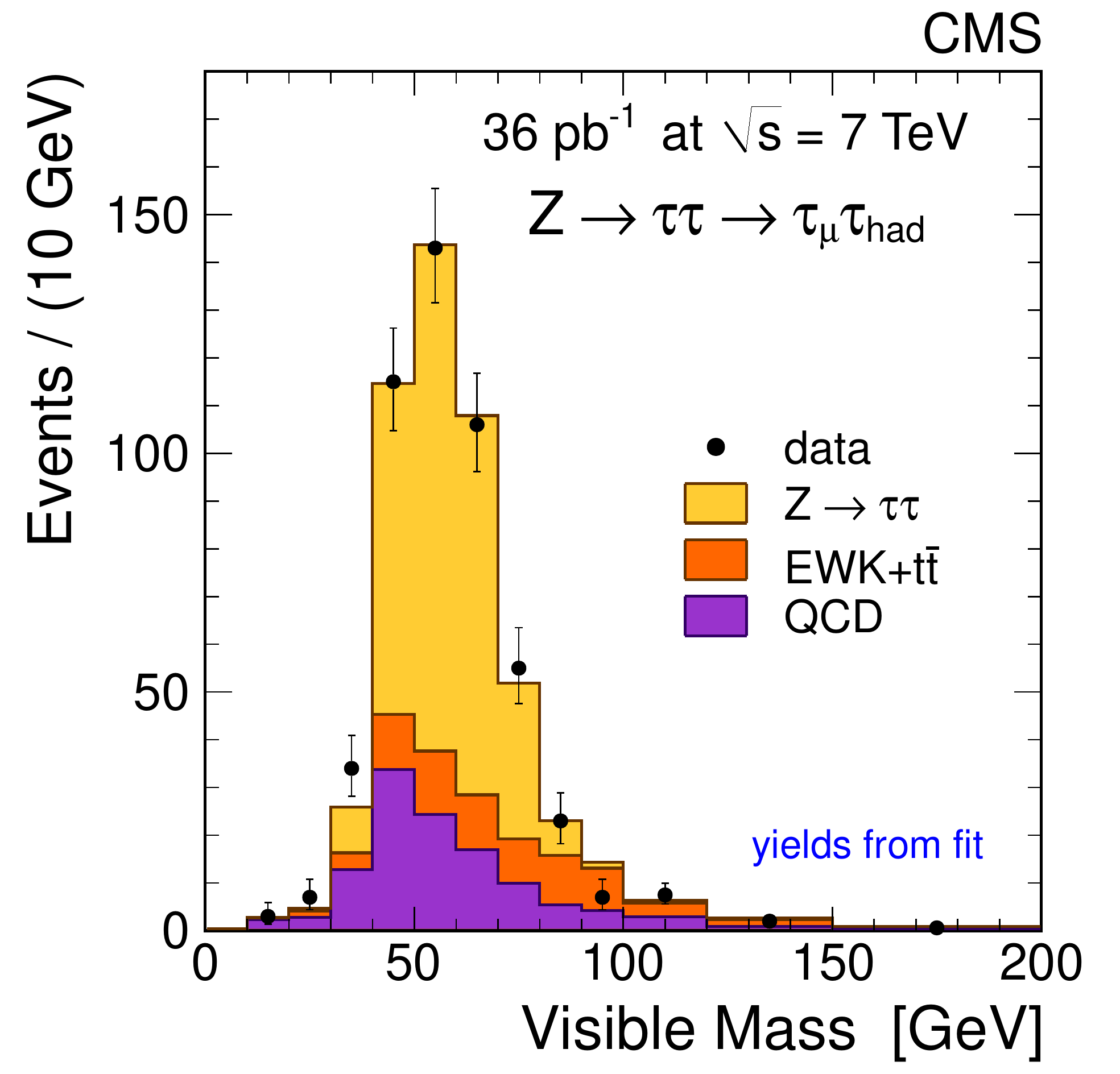}
\includegraphics[scale=0.35]{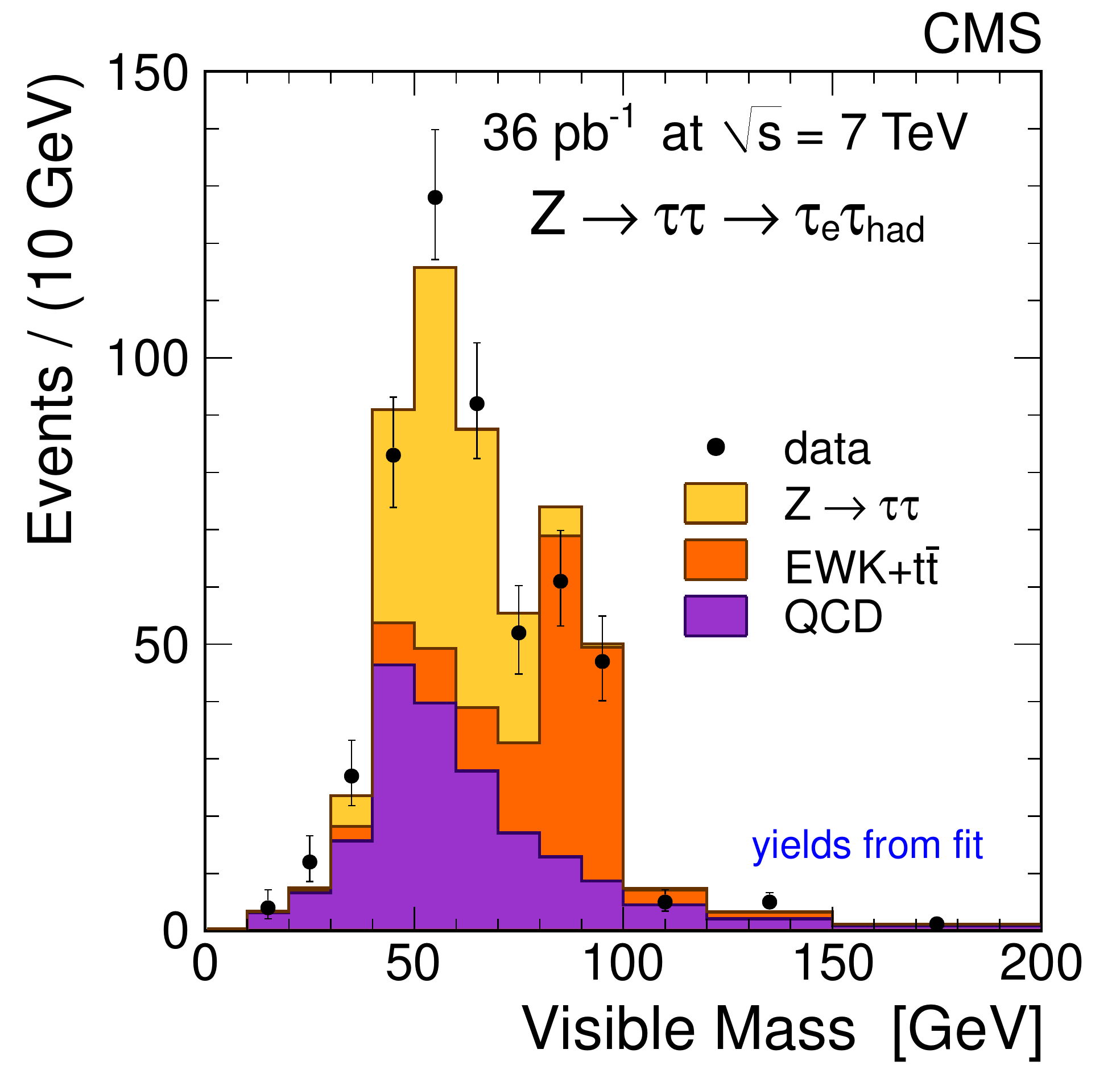}
\includegraphics[scale=0.35]{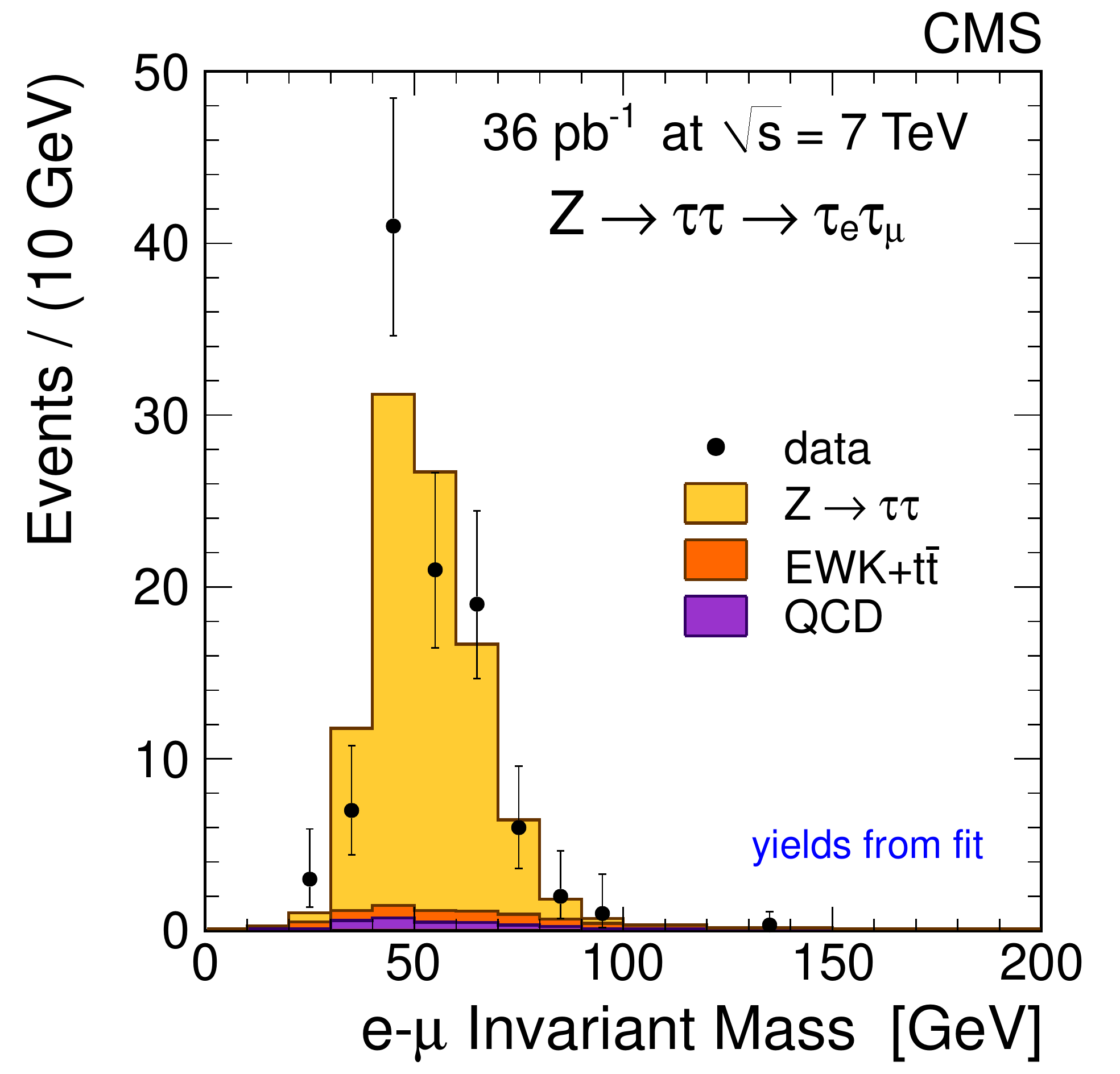}
\includegraphics[scale=0.35]{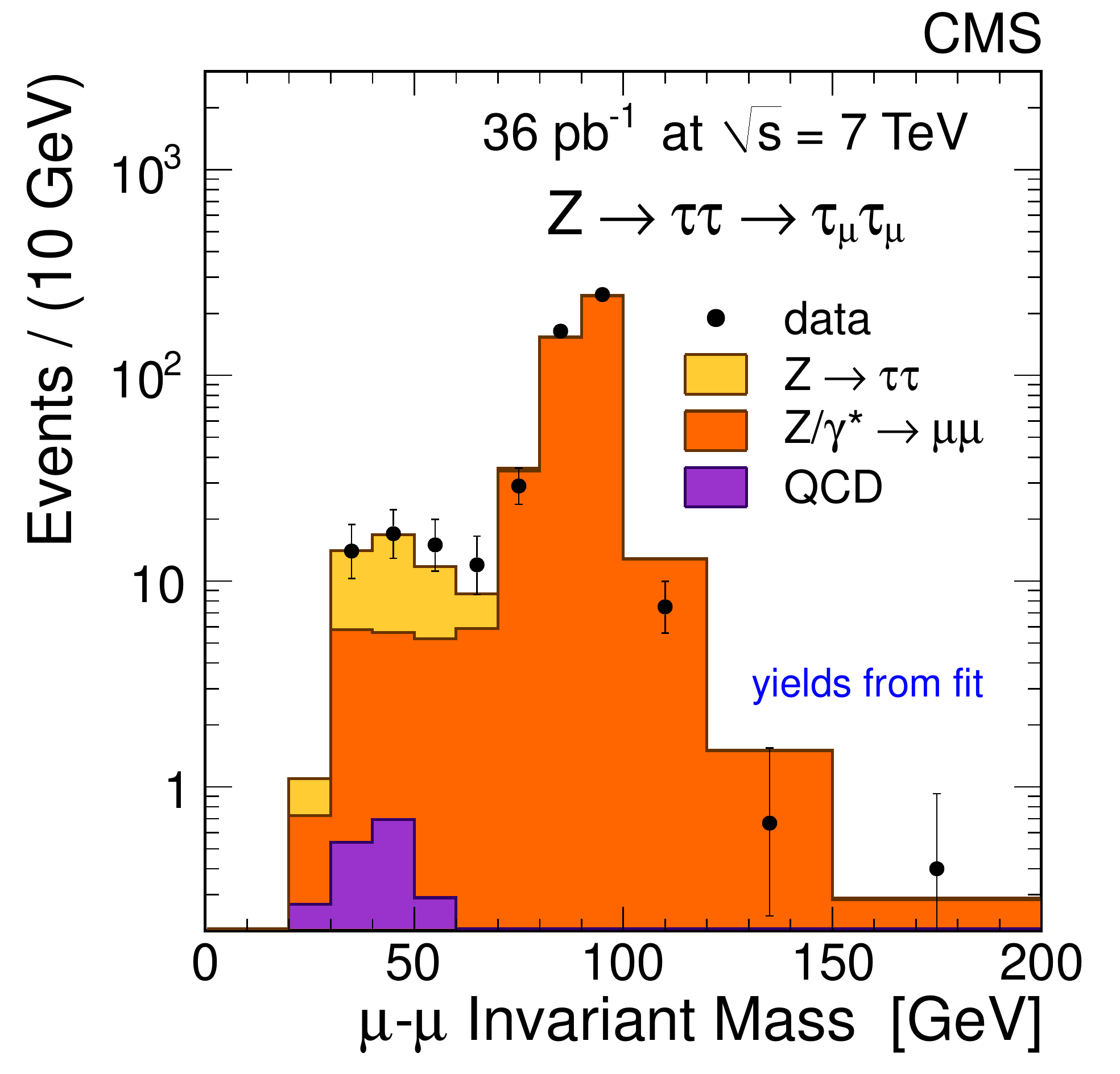}
\caption{Visible mass distributions of the $\taumu\tauhad$ (top left),
  $\taue\tauhad$ (top right), $\taue\taumu$ (bottom left), and
  $\taumu\taumu$ (bottom right) final states.}
\label{figure:inv_mass}
\end{center}
\end{figure}

The visible mass distributions of the $\taumu\tauhad$, $\taue\tauhad$, 
$\taue\taumu$ and $\taumu\taumu$  final states are shown in 
Fig.~\ref{figure:inv_mass}. 
To extract the signal, a fit is performed using the visible mass shapes
from the simulation, except for the QCD multijet and $\PZ  \rightarrow \ell^+\ell^-$ backgrounds, which are
obtained from data. For the simulation shapes, a variation of the electron and tau
energy scales within their uncertainties is considered. The effect of the muon energy scale is negligible. 
The background normalizations correspond to the numbers listed in 
Table~\ref{table:background_tot} and are allowed to vary within the estimated uncertainties. 
The background yields and signal shapes shown in Fig.~\ref{figure:inv_mass} are those obtained from the fitting procedure.

\begin{table}[!h]
\caption{Acceptance, selection efficiency and fraction of selected events outside the generator-level mass window for the four final states considered.} \vspace*{0.2cm}
\centering
\begin{tabular}{|l|c|c|c|c|}
\hline
 & $\taumu\tauhad$  & $\taue\tauhad$ &  $\taue\taumu$ &  $\taumu\taumu$  \\ 
\hline
\hline
Acceptance $\cal{A}$ & 0.13 & 0.12 & 0.074 & 0.16 \\
\hline
Selection efficiency $\epsilon$ & 0.37 & 0.23 & 0.55 & 0.17 \\
\hline
Mass window correction $f_{\rm out}$ & 0.03 & 0.03 & 0.02 & 0.01 \\
\hline
\end{tabular}
\label{table:acceptance}
\end{table}

The acceptances were obtained with the NLO QCD 
program {\sc POWHEG}
in the $\Ztautau$ mass region $60 < M_{\Pgt^+\Pgt^-} < 120$\GeV. 
Table~\ref{table:acceptance} shows the acceptances and the 
selection efficiencies for the different final states considered. 
The number of extracted events from the fit, $N_{\rm fit}$, is corrected for the fraction 
of signal events outside the generator-level mass window, $f_{\rm out}$,  
where $N = N_{\rm fit}\cdot (1-f_{\rm out})$ in Eq.~\ref{eq:crosssection}. The correction factors used
are also shown in Table~\ref{table:acceptance}.

The measured values of the cross section from the four final states considered are shown in Table~\ref{table:cross_section},
where the uncertainties shown are due to statistical, systematic, integrated luminosity and $\Pgt$ identification uncertainties.

\begin{table}[!h]
\caption{The measured values of the cross section from the four final states considered. The statistical, systematic and luminosity uncertainties are given. The uncertainty associated to the $\tauhad$ reconstruction and identification efficiency, $\tauhad-\mathrm{ID}$, is shown separately.}
\vspace*{0.2cm}
\centering
\begin{tabular}{|l|c|c|c|c|c|}
\hline
Final state & $\SIGBR{\PZ }{\Pgt^+\Pgt^-}$ nb & stat. & syst. & lumi. & $\Pgt\,\mathrm{ID}$ \\
\hline
$\taumu\tauhad$  & 0.83 & 0.07 & 0.04 & 0.03 & 0.19 \\
\hline
$\taue\tauhad$  & 0.94 & 0.11 & 0.03 & 0.04 & 0.22 \\
\hline
$\taue\taumu$ & 0.99 & 0.12 & 0.06 & 0.04 & \\
\hline
$\taumu\taumu$ & 1.14 & 0.27 & 0.04 & 0.05 & \\
\hline
\end{tabular}
\label{table:cross_section}
\end{table}

The measured values are compatible with each other and with the NNLO theoretical prediction, $0.972 \pm 0.042$~nb~\cite{Melnikov:2006kv}.
They are also consistent with the CMS measurement based on 
$\PZ \rightarrow \Pep\Pem$, $\Pgmp\Pgmm$ events~\cite{Khachatryan:2010xn}.

\par
\begin{figure}[!h]
\begin{center}
\includegraphics[scale=0.39]{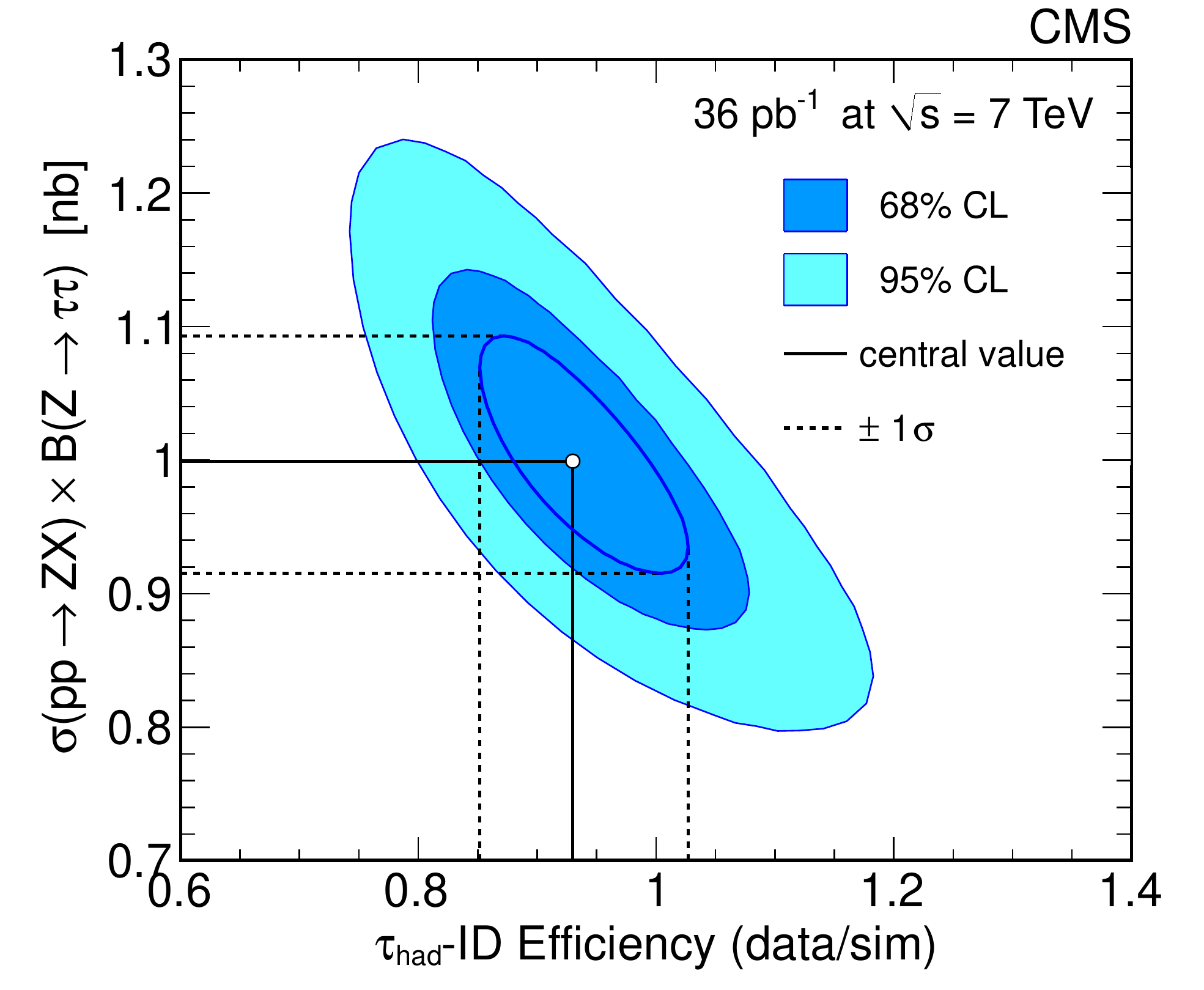}
\includegraphics[scale=0.39]{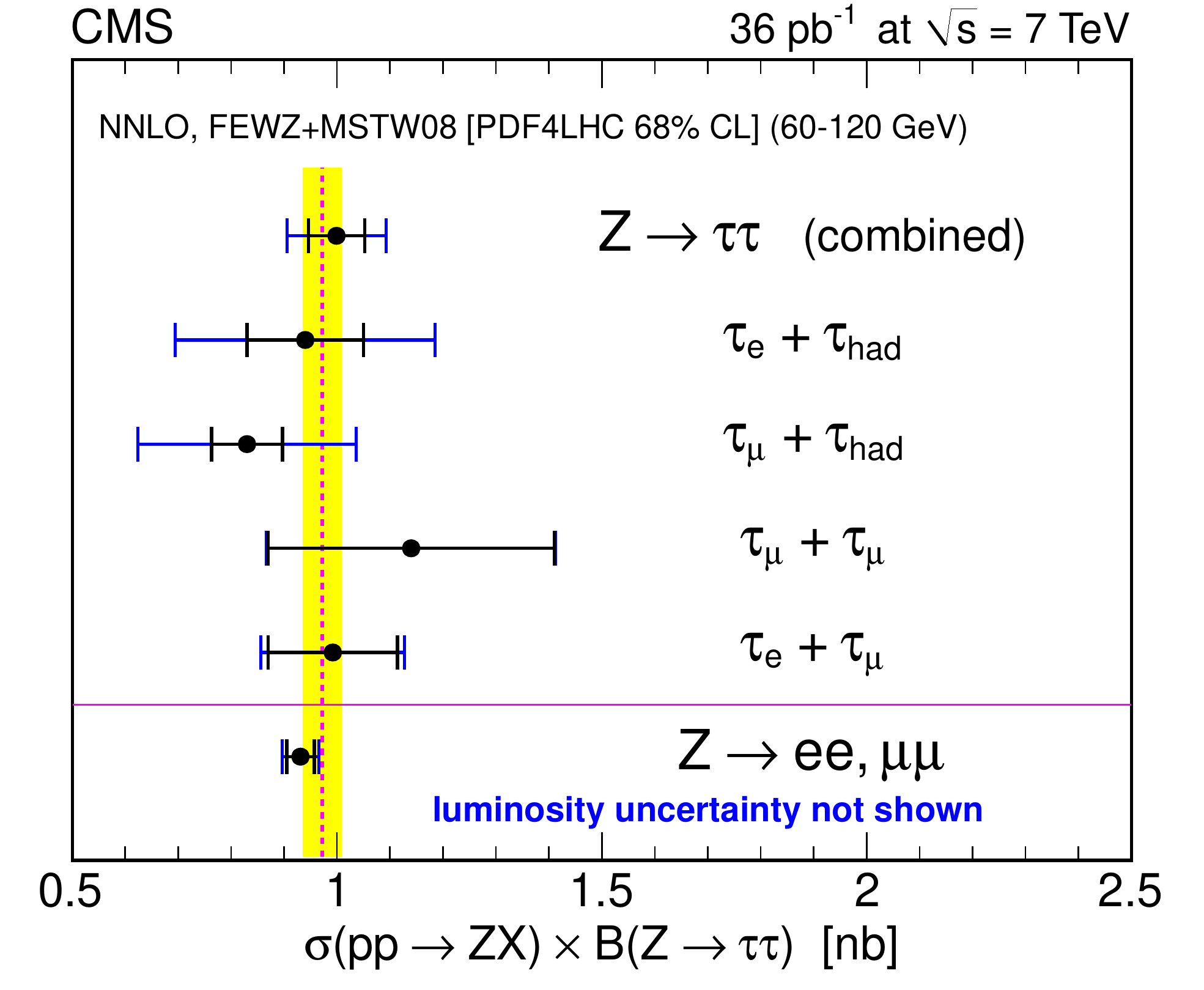}
\caption{Likelihood contours for the joint parameter estimation of the cross section and the $\Pgt$-identification (left). The fitted central values (solid line) and their estimated 1$\sigma$ uncertainties (dashed lines) are also shown.
Summary of the measured $\Ztautau$ cross sections in the 
$\taumu \tauhad$, $\taue \tauhad$, $\taue \taumu$,  
and $\taumu \taumu$ final states, in the invariant mass range of 
$60 < M_{\Pgt^+\Pgt^-} < 120$\GeV (right). The inner error bar represents the statistical uncertainty. 
The extracted cross section from the combined fit and the NNLO theoretical prediction 
are also shown.} 
\label{figure:contour}
\end{center}
\end{figure}

The dominant uncertainty on the $\Ztautau$ cross section 
measurement comes from the $\tauhad$ reconstruction and identification
efficiency.  
A simultaneous fit to all four
final states is performed to obtain the cross section and a scale factor for the $\tauhad$
efficiency, which is the ratio of the efficiency in the data to that in the simulation.  
The result of the global fit is 
shown in Fig.~\ref{figure:contour} (left), where
the likelihood contours for the best estimates of the cross section and the
$\tauhad$ efficiency scale factor are shown. In addition to the one-standard-deviation contour,  
the contours for which the likelihood $L$ is reduced by $2\Delta \ln L$ = 2.30 and 6.18 compared to the maximum value are also shown, corresponding to a coverage of $68\%$ and $95\%$ in the two-parameter plane, respectively.
The value of the cross section extracted from the fit is
\begin{displaymath}
\SIGBR{\PZ }{\Pgt^+\Pgt^-} = \RESA{1.00}{0.05}{0.08}{0.04},
\end{displaymath}
which is compared to the individual final state measurements 
in Fig.~\ref{figure:contour} (right). 
The value for the cross section is dominated by the dilepton
final states which have smaller systematic uncertainties. 
In the simultaneous fit, the $\tauhad$-ID scale factor is measured to be $0.93 \pm 0.09$.

A more precise value of the hadronic tau reconstruction efficiency can be obtained
by performing 
a fit of the $\taumu\tauhad$ and $\taue\tauhad$ final states, where the cross section 
is fixed to the value measured by CMS in the electron and muon decay channels,
\SIGBR{\PZ }{\Pep\Pem, \Pgmp\Pgmm} = \RESA{0.931}{0.026}{0.023}{0.102}~\cite{Khachatryan:2010xn}.
The extracted value of the $\tauhad$-ID scale factor is
$0.96 \pm 0.07$, which corresponds to a 
tau identification efficiency of $(47.4 \pm 3.3)\%$ in data,
for hadronically decaying tau leptons in the $\Ztautau$ sample with visible $\PT > 20$\GeV 
within the detector acceptance.

\section{Conclusions} \label{sec:conclusions}
\par
A measurement of the cross section for the 
process $\Pp\Pp\rightarrow \PZ X$ with $\Ztautau$ has been
performed, based on the $\taumu\tauhad$, $\taue\tauhad$,
$\taue\taumu$, and $\taumu\taumu$ final states.
A clear signal is established in the visible mass distributions 
for all channels.  The measured cross section, $\SIGBR{\mathrm{Z}}{\Pgt^+\Pgt^-} = \RESA{1.00}{0.05}{0.08}{0.04}$, 
is consistent 
with theoretical expectations, and the CMS measurement
for the $Z\rightarrow \Pep\Pem$ and $\Pgmp\Pgmm$ decay channels.  A global
fit of the $\taumu\tauhad$ and $\taue\tauhad$ channels 
provides a $7\%$ constraint on the efficiency for reconstructing
hadronic tau decays in CMS.

\section*{Acknowledgements \label{s:ack}}
We wish to congratulate our colleagues in the CERN accelerator departments for
the excellent performance of the LHC machine. We thank the technical and
administrative staff at CERN and other CMS institutes, and acknowledge support
from: FMSR (Austria); FNRS and FWO (Belgium); CNPq, CAPES, FAPERJ, and FAPESP
(Brazil); MES (Bulgaria); CERN; CAS, MoST, and NSFC (China); COLCIENCIAS
(Colombia); MSES (Croatia); RPF (Cyprus); Academy of Sciences and NICPB
(Estonia); Academy of Finland, ME, and HIP (Finland); CEA and CNRS/IN2P3
(France); BMBF, DFG, and HGF (Germany); GSRT (Greece); OTKA and NKTH (Hungary);
DAE and DST (India); IPM (Iran); SFI (Ireland); INFN (Italy); NRF (Korea); LAS
(Lithuania); CINVESTAV, CONACYT, SEP, and UASLP-FAI (Mexico); PAEC (Pakistan);
SCSR (Poland); FCT (Portugal); JINR (Armenia, Belarus, Georgia, Ukraine,
    Uzbekistan); MST and MAE (Russia); MSTDS (Serbia); MICINN and CPAN (Spain);
Swiss Funding Agencies (Switzerland); NSC (Taipei); TUBITAK and TAEK (Turkey);
STFC (United Kingdom); DOE and NSF (USA).
\bibliography{auto_generated}   

\cleardoublepage\appendix\section{The CMS Collaboration \label{app:collab}}\begin{sloppypar}\hyphenpenalty=5000\widowpenalty=500\clubpenalty=5000\textbf{Yerevan Physics Institute,  Yerevan,  Armenia}\\*[0pt]
S.~Chatrchyan, V.~Khachatryan, A.M.~Sirunyan, A.~Tumasyan
\vskip\cmsinstskip
\textbf{Institut f\"{u}r Hochenergiephysik der OeAW,  Wien,  Austria}\\*[0pt]
W.~Adam, T.~Bergauer, M.~Dragicevic, J.~Er\"{o}, C.~Fabjan, M.~Friedl, R.~Fr\"{u}hwirth, V.M.~Ghete, J.~Hammer\cmsAuthorMark{1}, S.~H\"{a}nsel, M.~Hoch, N.~H\"{o}rmann, J.~Hrubec, M.~Jeitler, G.~Kasieczka, W.~Kiesenhofer, M.~Krammer, D.~Liko, I.~Mikulec, M.~Pernicka, H.~Rohringer, R.~Sch\"{o}fbeck, J.~Strauss, F.~Teischinger, P.~Wagner, W.~Waltenberger, G.~Walzel, E.~Widl, C.-E.~Wulz
% ...
\vskip\cmsinstskip
% ...
1:~~Also at CERN, European Organization for Nuclear Research, Geneva, Switzerland\\
% ...

\end{sloppypar}
\end{document}